\documentclass[%
	reprint,
	superscriptaddress,
	amsmath,amssymb,
	aps,
	prl,
]%
{revtex4-2}

\usepackage[utf8]{inputenc}
\usepackage[T1]{fontenc}
\usepackage{lmodern}
\usepackage[english]{babel}
\usepackage[autostyle=true]{csquotes}

\usepackage{geometry}
\usepackage{hyperref}
\hypersetup{
		pdftitle={Dynamics of Stripe Patterns in Supersolid Spin-Orbit-Coupled Bose Gases},
		pdfauthor={Kevin T. Geier, Giovanni I. Martone, Philipp Hauke, Wolfgang Ketterle, Sandro Stringari},
}

\usepackage[dvipsnames]{xcolor}
\usepackage{graphicx}
\graphicspath{{figures/}}

\usepackage[%
	caption=false,
	subrefformat=simple,
	labelformat=simple,
]{subfig}

\usepackage{bm}
\usepackage{mathtools}
\usepackage{dsfont}
\usepackage{xspace}
\usepackage{xfrac}

\usepackage{mleftright}
\mleftright

\usepackage[%
    separate-uncertainty=false,
    range-phrase=-,
]{siunitx}

\DeclareSIUnit\bohr{\text {\ensuremath {a}}_{0}}

\usepackage[
	capitalise,
]{cleveref}

\usepackage{acro}
\acsetup{%
	first-style=long-short,
}

\DeclareAcronym{bec}{%
	short = BEC,
	long = Bose--Einstein condensate,
}

\DeclareAcronym{so}{%
	short = SO,
	long = spin--orbit,
}

\DeclareAcronym{soc}{%
	short = SOC,
	long = spin--orbit coupling,
}

\DeclareAcronym{socbec}{%
	short = SOC BEC,
	long = spin--orbit-coupled Bose--Einstein condensate,
}

\DeclareAcronym{gp}{%
	short = GP,
	long = Gross--Pitaevskii,
}

\DeclareAcronym{gpe}{%
	short = GPE,
	long = Gross--Pitaevskii equation,
}

\DeclareAcronym{1d}{%
	short = 1D,
	long = one-dimensional,
}

\DeclareAcronym{2d}{%
	short = 2D,
	long = two-dimensional,
}

\DeclareAcronym{3d}{%
	short = 3D,
	long = three-dimensional,
}

\DeclareAcronym{sm}{%
    short = SM,
    long = Supplemental Material,
}

\DeclareAcronym{tf}{%
    short = TF,
    long = Thomas--Fermi,
}

\newcommand*{\vect}[1]{\bm{#1}}
\newcommand*{\diff}{\mathop{}\!\mathrm{d}}

\DeclarePairedDelimiterX{\commutator}[2]{[}{]}{#1,#2}
\DeclarePairedDelimiterX{\anticommutator}[2]{\{}{\}}{#1,#2}

\DeclarePairedDelimiter{\abs}{\lvert}{\rvert}

\newcommand*{\transpose}{\intercal}

\DeclarePairedDelimiter{\braket}{\langle}{\rangle}
\DeclarePairedDelimiterX{\ketbra}[2]{|}{|}{#1\delimsize\rangle\delimsize\langle#2}

\renewcommand{\vec}{\vect}

\newcommand*{\recoilenergy}{\ensuremath{E_{r}}\xspace}
\newcommand*{\omegacritical}{\ensuremath{\Omega_{\mathrm{cr}}}\xspace}
\newcommand*{\upcomponent}{\ensuremath{\uparrow}\xspace}
\newcommand*{\downcomponent}{\ensuremath{\downarrow}\xspace}
\newcommand*{\densitycomponent}{\ensuremath{n}\xspace}
\newcommand*{\spincomponent}{\ensuremath{s}\xspace}
\newcommand*{\rubidium}{$^{87}$Rb\xspace}
\newcommand*{\potassium}{$^{39}$K\xspace}
\newcommand*{\lithium}{$^{7}$Li\xspace}

\begin{document}

\title{Dynamics of Stripe Patterns in Supersolid Spin--Orbit-Coupled Bose Gases}

\author{Kevin T. Geier}
\thanks{K.T.G.\ and G.I.M.\ contributed equally to this work.\\E-Mail K.T.G.\ at: \href{mailto:kevinthomas.geier@unitn.it}{kevinthomas.geier@unitn.it}\\E-Mail G.I.M.\ at: \href{mailto:giovanni_italo.martone@lkb.upmc.fr}{giovanni\_italo.martone@lkb.upmc.fr}}
\affiliation{Pitaevskii BEC Center, CNR-INO and Dipartimento di Fisica, Universit\`a di Trento, 38123 Trento, Italy}
\affiliation{Trento Institute for Fundamental Physics and Applications, INFN, 38123 Trento, Italy}
\affiliation{Institute for Theoretical Physics, Ruprecht-Karls-Universität Heidelberg, Philosophenweg 16, 69120 Heidelberg, Germany}

\author{Giovanni I. Martone}
\thanks{K.T.G.\ and G.I.M.\ contributed equally to this work.\\E-Mail K.T.G.\ at: \href{mailto:kevinthomas.geier@unitn.it}{kevinthomas.geier@unitn.it}\\E-Mail G.I.M.\ at: \href{mailto:giovanni_italo.martone@lkb.upmc.fr}{giovanni\_italo.martone@lkb.upmc.fr}}
\affiliation{Laboratoire Kastler Brossel, Sorbonne Universit\'{e}, CNRS, ENS-PSL Research University, Coll\`{e}ge de France; 4 Place Jussieu, 75005 Paris, France}
\affiliation{CNR NANOTEC, Institute of Nanotechnology, Via Monteroni, 73100 Lecce, Italy}
\affiliation{INFN, Sezione di Lecce, 73100 Lecce, Italy}

\author{Philipp Hauke}
\affiliation{Pitaevskii BEC Center, CNR-INO and Dipartimento di Fisica, Universit\`a di Trento, 38123 Trento, Italy}
\affiliation{Trento Institute for Fundamental Physics and Applications, INFN, 38123 Trento, Italy}

\author{Wolfgang Ketterle}
\affiliation{MIT-Harvard Center for Ultracold Atoms, Cambridge, Massachusetts 02138, USA}
\affiliation{Department of Physics, Massachusetts Institute of Technology, Cambridge, Massachusetts 02139, USA}

\author{Sandro Stringari}
\affiliation{Pitaevskii BEC Center, CNR-INO and Dipartimento di Fisica, Universit\`a di Trento, 38123 Trento, Italy}
\affiliation{Trento Institute for Fundamental Physics and Applications, INFN, 38123 Trento, Italy}

\date{\today}

\begin{abstract}
Despite ground-breaking observations of supersolidity in spin--orbit-coupled Bose--Einstein condensates, until now the dynamics of the emerging spatially periodic density modulations has been vastly unexplored.
Here, we demonstrate the nonrigidity of the density stripes in such a supersolid condensate and explore their dynamic behavior subject to spin perturbations.
We show both analytically in infinite systems and numerically in the presence of a harmonic trap how spin waves affect the supersolid's density profile in the form of crystal waves, inducing oscillations of the periodicity as well as the orientation of the fringes.
Both these features are well within reach of present-day experiments.
Our results show that this system is a paradigmatic supersolid, featuring superfluidity in conjunction with a fully dynamic crystalline structure.
\end{abstract}

%\keywords{}

\maketitle

\begin{figure}[t]
	\includegraphics[width=\columnwidth]{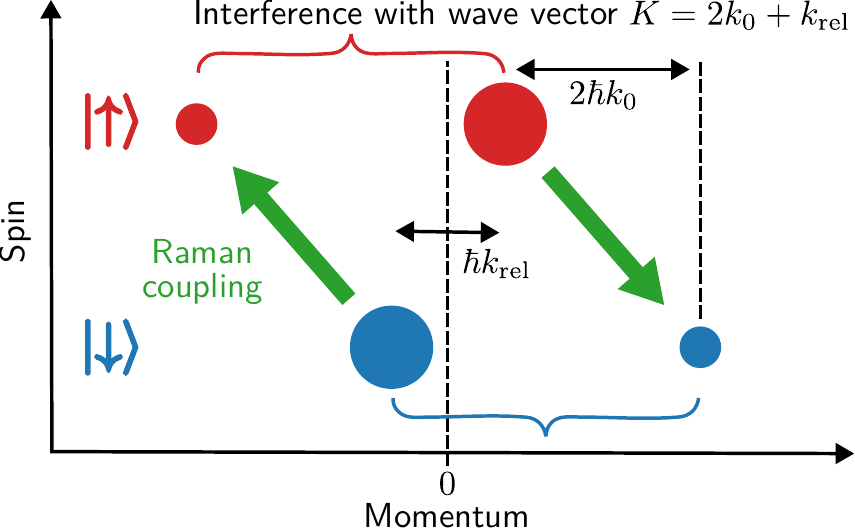}
	\caption{\label{fig:soc_stripes}%
		Illustration of the interference effects that lead to the appearance and dynamics of stripe patterns. The Raman process responsible for \acl{soc} turns the two-component \acl{bec} (two big circles) into a system with a four-component wave function. Components with the same spin form a spatial interference pattern.%
  }
\end{figure}

%%%%%%%%%%%%%%
% Introduction
%%%%%%%%%%%%%%

Supersolidity is an intriguing phenomenon exhibited by many-body systems, where both superfluid and crystalline properties coexist as a consequence of the simultaneous breaking of phase symmetry and translational invariance~\cite{Gross1957,Gross1958,Thouless1969,Andreev1969,Leggett1970,Kirzhnits1970}.
After unsuccessful attempts in solid helium~\cite{Balibar2010,Boninsegni2012}, supersolidity was first experimentally realized in \acp{bec} with \ac{soc}~\cite{Li2017,Putra2020} or inside optical resonators~\cite{Leonard2017}.
More recently, the supersolid phase has been identified in a series of experiments with dipolar Bose gases, where phase coherence, spatial modulations of the density profile, as well as the Goldstone modes associated with the superfluid and crystal behavior have been observed~\cite{Tanzi2019,Boettcher2019,Chomaz2019,Tanzi2019b,Guo2019,Natale2019,Petter2021,Chomaz2022}.

Since the experimental realization of \acp{socbec}~\cite{Lin2009,Lin2011}, this platform has emerged as a peculiar candidate of supersolidity because the spin degree of freedom is coupled to the density of the system~\cite{Wang2010,Ho2011,Wu2011,Li2012a,Li2013,Li2015review,Geier2021,Martone2021c}.
Without \ac{soc}, a two-component \ac{bec} has already two broken symmetries, one for the absolute phase and one for the relative phase between the two \ac{bec} order parameters. The addition of weak \ac{soc} mixes the spatial and spin degree of freedom, resulting in a stripe phase where the relative phase between the two condensates breaks the translational symmetry of space---the defining property of a supersolid. The Goldstone modes associated with the relative phase are spin excitations, whose dispersion relations as a function of the Raman coupling have been explored in Refs.~\cite{Chen2017,Geier2021}, but a connection to the crystal dynamics of the stripes has so far only been established for their rigid zero-frequency translational motion~\cite{Geier2021,Martone2021c}.

The rigidity of the stripe pattern has been controversially discussed in the literature.
For supersolids induced by coupling a \ac{bec} to two single-mode cavities~\cite{Leonard2017,Leonard2017a}, the wave vector of the density modulations is determined by the cavity light and the associated Goldstone mode is suppressed for nonzero wave vectors~\footnote{The situation is different for multi-mode cavities, see, e.g., Ref.~\cite{Guo2021}.}\hphantom{\vphantom{\cite{Guo2021}}}.
Since the spin--orbit effect is induced by Raman laser beams, it has been widely believed that the stripe pattern in \acp{socbec} is also externally imposed by the light and thus rigid.
Up to now, conclusive evidence for the nonrigidity of the stripe pattern has been lacking, as previous studies of stripe dynamics have mainly focused on the infinite-wavelength limit~\cite{Geier2021,Martone2021c}.
In this Letter, we elucidate the lattice-phonon nature of the spin Goldstone mode at finite wavelengths and thus demonstrate that the stripes form a fully dynamic crystal that is by no means rigid.
Specifically, we show how spin perturbations can excite oscillations of both the spacing and the orientation of the density fringes, establishing \acp{socbec} as paradigm examples of supersolidity.

%%%%%%%%%%%%%%%%%%%%%%%%%%%
% Origin of stripe dynamics
%%%%%%%%%%%%%%%%%%%%%%%%%%%

\textit{Origin of stripe dynamics.---}%
We consider the common scenario where \ac{soc} is generated in a binary mixture of atomic quantum gases by coupling two internal states using a pair of intersecting Raman lasers~\cite{Dalibard2011,Galitski2013,Aidelsburger2018}. In contrast to quantum mixtures with a simple coherent coupling of radio frequency or microwave type, the Raman coupling involves a finite momentum transfer $-2\hbar\vect{k}_0 = -2 \hbar k_0 \hat{\vect{e}}_x$, which we assume to point in the negative $x$~direction.
In the limit of weak Raman coupling, the emergence of the stripes and their dynamics can simply be explained as a spatial interference effect within a wave function which has four components (\cref{fig:soc_stripes}): spin-up condensate at zero momentum with a \ac{soc} admixture of spin down at wave vector $2 \vect{k}_0$, and spin-down condensate at zero momentum with a \ac{soc} admixture of spin up at wave vector $-2\vect{k}_0$. Because of \ac{soc}, there is now a spatial interference pattern between the two spin-up and two spin-down components with wave vector $\vect{K} = 2 \vect{k}_0$. The spontaneously chosen relative phase between the two condensates determines the origin of the stripe pattern. If there is a chemical potential difference between the two components, the relative phase of the two condensates will oscillate and therefore also the position of the stripes. One can add a spin current to the system, e.g., an out-of-phase or relative motion between the two condensates, which thus obtain the momenta $\pm \hbar \vect{k}_{\mathrm{rel}}/2$. The four components of the wave function are now at $\vect{k}_{\mathrm{rel}}/2$, $-\vect{k}_{\mathrm{rel}}/2-2\vect{k}_0$ for spin up and $-\vect{k}_{\mathrm{rel}}/2$, $\vect{k}_{\mathrm{rel}}/2+2\vect{k}_0$ for spin down, and the spatial interference pattern has now the wave vector $\vect{K} = 2\vect{k}_0 + \vect{k}_{\mathrm{rel}}$. If the spin current is oscillating, the wave vector of the stripe pattern will oscillate at the same frequency. When $\vect{k}_{\mathrm{rel}}$ is parallel to $\vect{k}_0$, the fringe spacing oscillates. When they are perpendicular, the angle of the fringes oscillates.
In what follows, we confirm and extend this intuitive picture using rigorous perturbative calculations and numerical simulations.

%%%%%%%%%%%%%%%%%%%%%%%
% Theoretical framework
%%%%%%%%%%%%%%%%%%%%%%%

\textit{Theoretical framework.---}%
After transforming to a spin-rotated frame, the single-particle Hamiltonian of the system takes the time-independent form~\cite{Lin2011}
\begin{equation}
\label{eq:hsoc}
    H_{\mathrm{SOC}} = \frac{1}{2m} \left( \vect{p} - \hbar \vect{k}_0 \sigma_z \right)^2 + \frac{\hbar \Omega}{2}\sigma_x + \frac{\hbar \delta}{2} \sigma_z + V(\vect{r}) \, ,
\end{equation}
where $m$ is the atomic mass, $\sigma_x$ and $\sigma_z$ are Pauli matrices, $\Omega$ is the strength of the Raman coupling, $\delta$ is the effective detuning, and $V(\vect{r})$ is a single-particle potential.
In infinite systems ($V \equiv 0$), the Hamiltonian is translationally invariant and allows for a spontaneous breaking of this symmetry, which, in combination with the broken $U(1)$ symmetry in the \ac{bec} phase, gives rise to supersolidity.

Since quantum depletion of a \ac{socbec} is typically small under realistic conditions~\cite{Zheng2013,Chen2018}, interactions between atoms are well described by mean-field theory via the \ac{gp} energy functional~\cite{Pitaevskii2016}
\begin{equation}
\label{eq:energyfunctional}
    E = \int \diff \vect{r} \, \left( \Psi^\dagger H_{\mathrm{SOC}} \Psi + \frac{g_{\densitycomponent \densitycomponent}}{2} n^2 + \frac{g_{\spincomponent \spincomponent}}{2} s_z^2 + g_{\densitycomponent \spincomponent} n s_z \right) \, .
\end{equation}
Here, the order parameter is given by a two-component spinor $\Psi = (\Psi_{\upcomponent}, \Psi_{\downcomponent})^\transpose$ with complex wave functions $\Psi_{\upcomponent}$ and $\Psi_{\downcomponent}$ for the individual spin states.
The last three terms in \cref{eq:energyfunctional} describe density--density, spin--spin, and density--spin interactions, respectively, where $n = \abs{\Psi_{\upcomponent}}^2 + \abs{\Psi_{\downcomponent}}^2$ denotes the total particle density and $s_z = \abs{\Psi_{\upcomponent}}^2 - \abs{\Psi_{\downcomponent}}^2$ is the spin density. The corresponding interaction constants $g_{\densitycomponent \densitycomponent} = (g_{\upcomponent \upcomponent} + g_{\downcomponent \downcomponent} + 2 g_{\upcomponent \downcomponent})/4$, $g_{\spincomponent \spincomponent} = (g_{\upcomponent \upcomponent} + g_{\downcomponent \downcomponent} - 2 g_{\upcomponent \downcomponent})/4$, and $g_{\densitycomponent \spincomponent} = (g_{\upcomponent \upcomponent} - g_{\downcomponent \downcomponent})/4$ are obtained from suitable combinations of the coupling constants $g_{ij} = 4 \pi \hbar^2 a_{ij} / m$, determined by the $s$-wave scattering lengths~$a_{ij}$ of the respective spin channels with $i,j \in \{ \upcomponent, \downcomponent \}$.
We focus our analysis on symmetric intraspecies interactions, assuming $g_{\densitycomponent \spincomponent} = 0$ and $\delta = 0$ from now on.

At the critical Raman coupling $\hbar \omegacritical = 4 \recoilenergy \sqrt{2 g_{\spincomponent \spincomponent} / (g_{\densitycomponent \densitycomponent} + 2 g_{\spincomponent \spincomponent})}$~\cite{Ho2011,Li2012a}, where $\recoilenergy = (\hbar k_0)^2 / 2 m$ is the recoil energy, the system undergoes a first-order transition between the supersolid (stripe) phase and the superfluid (but not supersolid) so-called plane-wave and single-minimum phases (see, e.g., Ref.~\cite{Li2015review}).
The latter are characterized by a strong Raman coupling that is responsible for the locking of the relative phase between the two spin components~\footnote{Note that the plane-wave phase leads to phase separation into two domains, one mainly spin up, the other mainly spin down. In each domain, the phase between spin-up and down components is locked.}, resulting from the competition between the spin ($g_{\spincomponent \spincomponent}$) and density ($g_{\densitycomponent \densitycomponent}$) interaction components of the mean-field energy functional~\labelcref{eq:energyfunctional}~\cite{Martone2012}.
Consequently, there is only a single spin--density-hybridized Goldstone mode above~$\omegacritical$.
Conversely, in the supersolid phase, below~$\omegacritical$, the spontaneous breaking of both phase and translational symmetry implies the existence of two Goldstone modes of predominantly density and spin nature with distinct sound velocities (see \ac{sm}~\cite{SM}\hphantom{\vphantom{\cite{Li2012}}} for further details)~\cite{Li2013,Martone2021c}.

A major question to be addressed in what follows is how the spin degree of freedom can induce dynamics in the stripe patterns and in particular how the excitation of a spin wave results in the excitation of a crystal wave affecting the time dependence of the density profile.

%%%%%%%%%%%%%%%%%%%%%%%%%%%%%%%%%%%%%%%%%%%
% Perturbation approach in infinite systems
%%%%%%%%%%%%%%%%%%%%%%%%%%%%%%%%%%%%%%%%%%%

\textit{Perturbation approach in infinite systems.---}%
A useful scenario to probe this question consists in suddenly releasing at time $t=0$ a small static spin perturbation of the form $- \lambda \recoilenergy \sigma_z
\cos(\vect{q} \cdot \vect{r})$, with $0 < \lambda \ll 1$. The wave vector~$\vect{q}$ is assumed to be small in order to explore the relevant phonon
regime, where a major effect of the release of the perturbation is the creation of a spin wave propagating with velocity~$c_\spincomponent$. Here, we are mainly
interested in its effect on the dynamic behavior of the stripes characterizing the density distribution. 
Starting from the results of Ref.~\cite{Martone2021c}
for the Bogoliubov amplitudes of the phonon modes in the long-wavelength limit, and neglecting the small
contributions of the gapped modes of the Bogoliubov spectrum, the space and time dependence of the density can be
written in the form
\begin{equation}
n(\vect{r},t) = \bar{n} + \sum_{\bar{m} = 1}^{+\infty} \tilde{n}_{\bar{m}} \cos[\bar{m} \chi(\vect{r},t)] \, .
\label{eq:tot_dens}
\end{equation}
Here, $\bar{n}$ is the average density and
\begin{equation}
\chi(\vect{r},t) = 2 k_1 x + \phi + \delta\phi(t) \cos(\vect{q} \cdot \vect{r})
\label{eq:rel_phase}
\end{equation}
the relative phase between the two condensates in the spin-rotated frame. The sum over the integer index~$\bar{m}$ reflects the presence of higher harmonics in the density profile~\eqref{eq:tot_dens} characterizing the stripe phase, whose coefficients are denoted by $\tilde{n}_{\bar{m}}$. \Cref{eq:tot_dens,eq:rel_phase} explicitly reveal that the perturbed density fringes are a combined effect of the equilibrium modulations, fixed by the wave vector $2 \vect{k}_1 = 2 k_1 \hat{\vect{e}}_x$ (which differs from $2 \vec{k}_0$ at finite Raman coupling~\cite{Li2012a,Martone2021c}), and those induced by the external perturbation, characterized by the wave vector~$\vect{q}$.
The perturbative expression for $k_1$ is reported in Ref.~\cite{Martone2021c} and for convenience in the \ac{sm}~\cite{SM}.
The phase~$\phi$ represents the spontaneously chosen offset of the stripe pattern in equilibrium. The time dependence of the function~$\delta\phi$ is fixed by
the sound velocities~$c_{\densitycomponent,\spincomponent}$ of the density and spin phonons as well as by the Raman coupling~$\Omega$. 

For $q \ll k_1$, the relative phase~\eqref{eq:rel_phase} varies very slowly over a large number of equilibrium density
oscillations. Consequently, in a region of space $|\vect{r} - \vect{r}_0| \ll q^{-1}$ around a given point $\vect{r}_0$, one can approximate $\chi$ by
its first-order Taylor expansion, 
\begin{equation}
\chi(\vect{r},t) \simeq \vect{K}(\vect{r}_0,t) \cdot \vect{r} + \Phi(\vect{r}_0,t) \, .
\label{eq:rel_phase_exp}
\end{equation}
This expression features a local time-dependent stripe wave vector, whose structure
\begin{equation}
\vect{K}(\vect{r}_0,t) = \nabla\chi(\vect{r}_0,t) = 2 \vec{k}_1 - \delta\phi(t) \sin(\vect{q} \cdot \vect{r}_0) \vec{q} 
\label{eq:loc_wave_vec}
\end{equation}
confirms the intuitive scenario of  \cref{fig:soc_stripes} (where $\vect{k}_1$ has been approximated by $\vect{k}_0$), upon identifying $\vect{k}_{\mathrm{rel}}$ with the second term in \cref{eq:loc_wave_vec}.
In addition, Eq.~\eqref{eq:rel_phase_exp} contains the phase shift
\begin{equation}
\begin{split}
\Phi(\vect{r}_0,t)
&{} = \chi(\vect{r}_0,t) - \vec{r}_0 \cdot \nabla\chi(\vect{r}_0,t) \\
&{} = \phi + \delta\phi(t) \left[ \cos(\vect{q} \cdot \vect{r}_0) + (\vect{q} \cdot \vect{r}_0) \sin(\vect{q} \cdot \vect{r}_0)\right] \, ,
\end{split}
\label{eq:loc_phase_shift}
\end{equation}
which is responsible for the time modulation of the offset of the stripe pattern.

Carrying out a perturbative analysis of the order parameter of the condensate up to second order in $\hbar\Omega/4\recoilenergy$ (see Refs.~\cite{Martone2021c,Martone2023prep}) yields the result
\begin{equation}
\delta\phi(t) = - \frac{2 k_1 v_\spincomponent}{c_\spincomponent q} \sin (c_\spincomponent q t) \, ,
\label{eq:pattern_mod}
\end{equation}
where we have introduced the velocity
\begin{equation}
v_\spincomponent = \lambda \frac{\hbar k_0}{m} \left[ \frac{1}{2} - \beta \left( \frac{\hbar\Omega}{4 \recoilenergy} \right)^2 \right] \, ,
\label{eq:spin_vel}
\end{equation}
with $\beta = \recoilenergy \bar{n} [2 \recoilenergy g_{\densitycomponent \densitycomponent}
+ 2(2 \recoilenergy + \bar{n} g_{\densitycomponent \densitycomponent}) g_{\spincomponent \spincomponent}
+ \bar{n} g_{\spincomponent \spincomponent}^2] / [2 (2 \recoilenergy + \bar{n} g_{\densitycomponent \densitycomponent})^2
(2 \recoilenergy + \bar{n} g_{\spincomponent \spincomponent})]$, and the expression for $c_\spincomponent$ is reported in Ref.~\cite{Martone2021c} and for convenience in the \ac{sm}~\cite{SM}.
At the leading order~$\Omega^2$, only the spin sound velocity~$c_\spincomponent$ enters \cref{eq:pattern_mod}, while a second term oscillating at the density phonon frequency~$c_\densitycomponent q$ appears at order $\Omega^4$~\cite{Martone2023prep}.
For $\Omega = 0$, the velocity~$v_\spincomponent$ fixes the time variation rate of the relative phase of the quantum mixture, without any consequence for the density distribution since the contrast of fringes exactly vanishes in this limit~\cite{Li2012a,Martone2021c} (see also \ac{sm}~\cite{SM}).

If $\cos(\vect{q} \cdot \vect{r}_0) = \pm 1$, the initial static spin perturbation has a peak (antinode) at $\vec{r}_0$, and close to this point it becomes of the form
$ \mp \lambda \recoilenergy \sigma_z$. After releasing the spin perturbation, there is a spin imbalance at $\vect{r}_0$ and the difference in chemical
potentials causes an oscillation of the relative phase of the two condensates.
From \cref{eq:loc_phase_shift,eq:pattern_mod} one sees that, at times satisfying the condition $t \ll (c_\spincomponent q)^{-1}$ (which is easily fulfilled for the small $q$ of interest here), the stripes show a displacement at velocity~$\pm v_\spincomponent$, i.e., $\chi(\vec{r},t) \simeq 2 k_1 (x \mp v_\spincomponent t) + \phi$, in excellent agreement with the numerical findings of Ref.~\cite{Geier2021} (see \ac{sm}~\cite{SM} for further details).
At $q=0$, the spatial translation of stripes corresponds to the zero-frequency limit of the spin Goldstone branch.

Far from the antinodes, after the spin quench there is an oscillating spin current, which makes also the stripe wave vector~\eqref{eq:loc_wave_vec} vary in time.
The strongest oscillations occur when $\sin(\vect{q} \cdot \vect{r}_0) = \pm 1$, i.e., $\vect{r}_0$ is a node of the initial perturbation,
which is thus antisymmetric under inversion with respect to $\vect{r}_0$ and locally behaves as $\pm \lambda \recoilenergy \vect{q} \cdot (\vect{r} - \vect{r}_0) \sigma_z$.
In particular, if $\vect{q} = q \hat{\vect{e}}_x$, the local stripe wavelength $2\pi/\abs{\vect{K}(\vect{r}_0,t)} =
[1 \mp (v_\spincomponent/c_\spincomponent) \sin (c_\spincomponent q t)] \pi/k_1$ oscillates around its equilibrium value.
By contrast, if $\vect{q} = q \hat{\vect{e}}_y$, the stripes rotate by an angle $\pm (v_\spincomponent/c_\spincomponent) \sin (c_\spincomponent q t)$ about the $z$~axis.
This effect occurs in combination with the fringe displacement seen above, unless $\vect{r}_0$ coincides with a maximum or minimum of the equilibrium density distribution.

The above discussion shows that a spin perturbation applied to the stripe configuration can cause a rigid motion of the stripes as well as a periodic change in either magnitude or orientation of their wave vector, depending on the local behavior of the perturbation.
Although the analytic results~\labelcref{eq:pattern_mod,eq:spin_vel} have been derived by carrying out a perturbative analysis up to order $\Omega^2$, we have verified that they provide a rather accurate description of the dynamics of stripes, as compared to a numerical solution of the time-dependent linearized \ac{gp} equation in infinite systems, also for fairly large values of the Raman coupling.

%%%%%%%%%%%%%%%%%%%%%%%%%%%%%%%%%%%%%%%%%%
% Numerical simulations in a harmonic trap
%%%%%%%%%%%%%%%%%%%%%%%%%%%%%%%%%%%%%%%%%%

\begin{figure}[t]
	\subfloat{\label{fig:compressionrotation:a}}%
	\subfloat{\label{fig:compressionrotation:b}}%
	\subfloat{\label{fig:compressionrotation:c}}%
	\subfloat{\label{fig:compressionrotation:d}}%
	\includegraphics[width=\columnwidth]{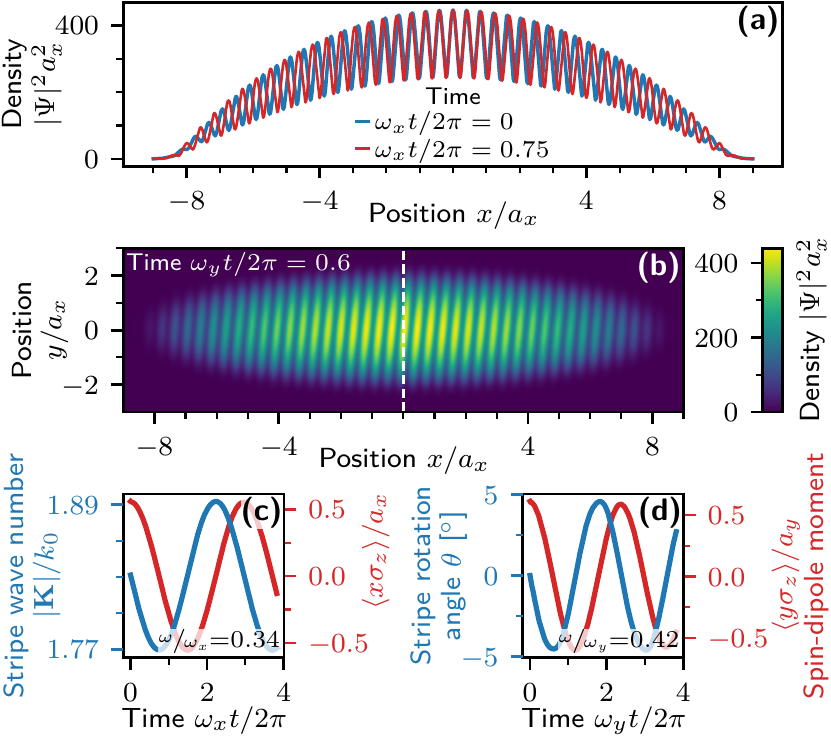}
	\caption{\label{fig:compressionrotation}%
        Dynamics of the stripe pattern in a harmonically trapped system for $\hbar \Omega / \recoilenergy = \num{1.75}$.
		(a),(b)~Snapshots of the density profile at different times, showing the compression and dilatation of the fringe spacing~(a) as well as the rotation of the stripes~(b) after suddenly releasing the longitudinal and transversal spin perturbations $H_{x \sigma_z} = -m \omega_x^2 x_0 \mathinner{x \sigma_z}$ with $x_0 / a_x = 0.1$ and $H_{y \sigma_z} = -m \omega_y^2 y_0 \mathinner{y \sigma_z}$ with $y_0 / a_y = 0.15$, respectively.
        (c)~Evolution of the magnitude of the stripe wave vector~$\abs{\vect{K}}$ and of the longitudinal spin-dipole moment~$\braket{x \sigma_z}$ for the scenario in~(a).
        (d)~Time trace of the rotation angle~$\theta$ of the stripes and of the transversal spin-dipole moment~$\braket{y \sigma_z}$ for the scenario in~(b).
        The oscillation frequencies of the stripe pattern coincide with those of the corresponding spin-dipole moments.%
	}
\end{figure}

\textit{Numerical simulations in a harmonic trap.---}%
Having understood how spin perturbations affect the dynamics of the stripe pattern in infinite systems, we now illustrate similar effects taking place in finite-size configurations, namely, in the presence of a harmonic trapping potential $V(\vect{r}) = m (\omega_x^2 x^2 + \omega_y^2 y^2 + \omega_z^2 z^2) / 2$  with angular frequencies~$\omega_i$ and corresponding oscillator lengths $a_i = \sqrt{\hbar / m \omega_i}$, $i = x, y, z$.
To this end, we numerically solve the full time-dependent \ac{gp} equations, which can be derived by applying the variational principle $i \hbar \partial_t \Psi_{\upcomponent / \downcomponent} = \delta E / \delta \Psi_{\upcomponent / \downcomponent}^*$ to the energy functional~\labelcref{eq:energyfunctional}.

For our numerics, we assume symmetric intraspecies scattering lengths close to those of \rubidium,  where the majority of experiments on \acp{socbec} has been conducted, $a_{\upcomponent \upcomponent} = a_{\downcomponent \downcomponent} = \SI{100}{\bohr}$ ($\si{\bohr}$ is the Bohr radius).
Moreover, to enhance the miscibility of the two spin species and thus the supersolid features, we assume a quasi-\ac{2d} situation with reduced interspecies coupling $\tilde{g}_{\upcomponent \downcomponent} = \num{0.6} \, \tilde{g}_{\upcomponent \upcomponent}$ (see experimental considerations below), where $\tilde{g}_{\upcomponent \upcomponent} = \tilde{g}_{\downcomponent \downcomponent} = g_{\upcomponent \upcomponent} / \sqrt{2 \pi} a_z$ are effective \ac{2d} couplings for a strong vertical confinement with frequency $\omega_z / 2 \pi = \SI{2500}{\hertz}$~\cite{Martone2014}.
Further, we choose an elongated trap in the $x$~direction with $(\omega_x, \omega_y) = 2 \pi \, (\num{50}, \num{200}) \, \si{\hertz}$, a total particle number of $N = \num{e4}$, as well as $k_0 = \sqrt{2} \pi / \lambda_0$ with $\lambda_0 = \SI{804.1}{\nano\meter}$~\cite{Lin2011}.

The numerical protocol is the same as that in the quench scenario considered above: we first compute the ground state in the presence of a small static perturbation of spin nature and then observe the dynamics after suddenly releasing the perturbation at time $t = 0$.
Here, our analysis is focused on the stripe dynamics generated by the longitudinal and transversal spin operators $x \sigma_z$ and $y \sigma_z$, which correspond in infinite systems to the local behavior of the perturbation around the nodes.
The translational motion of the stripes induced by the uniform spin operator~$\sigma_z$, corresponding to a sudden change of the Raman detuning, has been studied numerically in Ref.~\cite{Geier2021} and is further detailed in the \ac{sm}~\cite{SM}.

\Cref{fig:compressionrotation:a,fig:compressionrotation:b} illustrate, respectively, the oscillation of the fringe spacing and the periodic rotation of the stripe wave vector in response to weak perturbations by the operators $x \sigma_z$ and $y\sigma_z$.
The corresponding oscillation frequencies coincide with those of the induced spin-dipole oscillations $\braket{x \sigma_z}$ and $\braket{y \sigma_z}$, as shown in \cref{fig:compressionrotation:c,fig:compressionrotation:d}, respectively.
It is remarkable that the transversal spin operator~$y\sigma_z$, which generates the oscillating rotation of the stripes in the supersolid phase, also constitutes a crucial spin contribution to the angular momentum operator as a consequence of \ac{soc}~\cite{Radic2011,Qu2018}.
The inclusion of such an effective $y$-dependent detuning has been used to generate quantized vortices~\cite{Lin2009} and to show the occurrence of crucial rigid components in the moment of inertia~\cite{Stringari2017}.

Unsurprisingly, owing to nonlinear effects in the \ac{soc} strength, the dynamic excitation of stripes is not only produced by spin perturbations (as considered above), but also by density perturbations.
A density perturbation mainly excites the associated density Goldstone mode, but due to \ac{soc} also produces a weak cross excitation of the spin Goldstone mode.
Since the density mode also has a weak manifestation in the spin sector, quantities sensitive to the spin degree of freedom, e.g., the fringes, exhibit a beat note involving the frequencies of both Goldstone modes~\footnote{Vice versa, a spin perturbation also generates a beat note in density observables.}.
In fact, one can show that a density perturbation generates a beating oscillation of the stripe wave vector with an amplitude of order~$\Omega^2$~\cite{Martone2023prep} (see \ac{sm}~\cite{SM} for an illustration of such beating effects in harmonically trapped systems).
By contrast, a spin perturbation produces a strong excitation of the spin mode (and thus of the stripe pattern) with practically invisible beating in the stripe wave vector since the contribution of the density mode is of order~$\Omega^4$, as noted below \cref{eq:spin_vel}.

%%%%%%%%%%%%%%%%%%%%%%%%%%%
% Experimental perspectives
%%%%%%%%%%%%%%%%%%%%%%%%%%%

\textit{Experimental perspectives.---}%
For the study of stripe dynamics, it is favorable to have stripes with high contrast.  This requires strong miscibility between the two components to suppress the transition to the phase-separated plane-wave phase.  It is therefore best to use an atom where the scattering lengths are tunable via Feshbach resonances, such as \potassium~\cite{Jorgensen2016} or \lithium~\cite{Secker2020}.
Alternatively, in species with low bulk miscibility, such as \rubidium, the critical Raman coupling for the stripe phase can be enhanced by considering a quasi-\ac{2d} configuration characterized by a reduced spatial overlap of the two spin components in the strongly confined direction.
This can be realized experimentally with the help of a spin-dependent trapping potential~\cite{Martone2014,Martone2015,Hond2022} (as we have assumed in our numerics above) or using pseudospin orbital states in a superlattice~\cite{Li2017}.

The stripe pattern for \acp{socbec} has been observed via Bragg scattering~\cite{Li2017,Putra2020}. Since the Bragg angle depends on the period (and angle) of the stripes, any oscillation in the stripe spacing (or orientation) will result in an oscillating Bragg signal.
Our simulations for realistic parameters show a modulation of the stripe wave vector on the order of $\SI{5}{\percent}$. This should be easily resolvable in experiments since the angular resolution of the Bragg spot is diffraction limited by the condensate size, which is typically 10 to 50 times larger than the fringe spacing.
Alternatively, the periodic dilatation or rotation of the stripe wave vector could be observed by identifying the oscillating peaks in the momentum distribution after ballistic expansion.
The dynamics of the stripes, including their zero-frequency translational motion, may also be observed \textit{in situ} after increasing the stripe period to several microns, e.g., by creating a spatial beat note with the pattern imprinted by a $\pi/2$ Raman pulse~\cite{Martone2014} or by using matter-wave-lensing techniques~\cite{Murthy2014,Tarruell2022}.
Interestingly, since the phase of the Raman beams is added to the spontaneous phase due to symmetry breaking, an oscillation of the position of the stripes can be driven by a frequency detuning of the Raman beams and could possibly be detected by an increase in temperature after dissipative damping.

%%%%%%%%%%%%
% Conclusion
%%%%%%%%%%%%

In conclusion, \ac{soc} supersolids display a rich dynamics of their spontaneously established crystal order.  This is similar to the dynamics predicted and observed in dipolar quantum gases~\cite{Tanzi2019b,Guo2019,Natale2019}.
The main difference between the two systems is that \ac{soc} supersolids have a spin degree of freedom, which provides a natural way to excite the crystal Goldstone mode.
This supersolid Goldstone mode is of hybridized spin--density nature.
Its dynamics is different from that of supersolids mediated by two single-mode cavities, where nonzero wave vectors are suppressed by the infinite-range coupling~\cite{Leonard2017,Leonard2017a}, and in strong distinction from externally imposed rigid density patterns as in optical lattices.
As we have shown, the predicted dynamics in \ac{soc} supersolids is readily accessible within state-of-the-art experimental capabilities.

% \appendix

%%%%%%%%%%%%%%%%%
% Acknowledgments
%%%%%%%%%%%%%%%%%

\begin{acknowledgments}
This project has received funding from the European Research Council (ERC) under the European Union’s Horizon 2020 research and innovation programme (Grant Agreement No.\ 804305).
This work has been supported by Q@TN, the joint lab between the University of Trento, FBK---Fondazione Bruno Kessler, INFN---National Institute for Nuclear Physics, and CNR---National Research Council.
We further acknowledge support by Provincia Autonoma di Trento, from the Italian Ministry of University and Research (MUR) through the PRIN project INPhoPOL (Grant No.\ 2017P9FJBS) and the PNRR MUR Project No.\ PE0000023---NQSTI, and from the National Science Foundation through the Center for Ultracold Atoms and Grant No.\ 1506369.
\end{acknowledgments}


%%%%%%%%%%%%%%%%%%%%%%%
% Supplemental Material
%%%%%%%%%%%%%%%%%%%%%%%

\cleardoublepage

\acresetall

\setcounter{equation}{0}
\setcounter{figure}{0}
\setcounter{table}{0}
\setcounter{page}{1}
\setcounter{section}{0}
\setcounter{secnumdepth}{2}

\makeatletter
\renewcommand{\theequation}{S\arabic{equation}}
\renewcommand{\thefigure}{S\arabic{figure}}
\renewcommand{\thetable}{S\arabic{table}}
\renewcommand{\bibnumfmt}[1]{[S#1]}
\renewcommand{\citenumfont}[1]{S#1}
\makeatother

\onecolumngrid

\begin{center}
	\textbf{\large%
		Supplemental Material:\\Dynamics of Stripe Patterns in Supersolid Spin--Orbit-Coupled Bose Gases%
	}\\[1em]
	{\normalfont\normalsize%
		Kevin T. Geier,$^{1,2,3,*}$ Giovanni I. Martone,$^{4,5,6,*}$ Philipp Hauke,$^{1,2}$ Wolfgang Ketterle,$^{7,8}$ and Sandro Stringari$^{1,2}$%
	}\\[1ex]
	\textit{\small%
		${}^1$Pitaevskii BEC Center, CNR-INO and Dipartimento di Fisica, Universit\`a di Trento, 38123 Trento, Italy\\
		${}^2$Trento Institute for Fundamental Physics and Applications, INFN, 38123 Trento, Italy\\
		${}^3$Institute for Theoretical Physics, Ruprecht-Karls-Universität Heidelberg, Philosophenweg 16, 69120 Heidelberg, Germany\\
		${}^4$Laboratoire Kastler Brossel, Sorbonne Universit\'{e}, CNRS,\\
		ENS-PSL Research University, Coll\`{e}ge de France; 4 Place Jussieu, 75005 Paris, France\\
		${}^5$CNR NANOTEC, Institute of Nanotechnology, Via Monteroni, 73100 Lecce, Italy\\
		${}^6$INFN, Sezione di Lecce, 73100 Lecce, Italy\\
		${}^7$MIT-Harvard Center for Ultracold Atoms, Cambridge, Massachusetts 02138, USA\\
		${}^8$Department of Physics, Massachusetts Institute of Technology, Cambridge, Massachusetts 02139, USA%
	}\\
	{\normalfont\small%
		(Dated: \today)%
	}
\end{center}

\vspace{1ex}

\begingroup
\addtolength\leftmargini{3em}
\small
\begin{quotation}
In this Supplemental Material, we provide further information about the dynamics of density fringes in a supersolid spin--orbit-coupled Bose--Einstein condensate.
These include the dispersion of collective oscillations (\cref{sec:disp_coll_osc}), details on the perturbation approach in infinite systems (\cref{sec:pert_app}), the translational motion of stripes following the release of a uniform spin perturbation (\cref{sec:stripe_mot}), and the beating effect caused by the release of a density perturbation (\cref{sec:beat_dens}).
\end{quotation}
\endgroup

\vspace{1em}

\thispagestyle{empty}

\twocolumngrid

\section{Dispersion of collective excitations}
\label{sec:disp_coll_osc}

\begin{figure}[t]
	\includegraphics[width=\columnwidth]{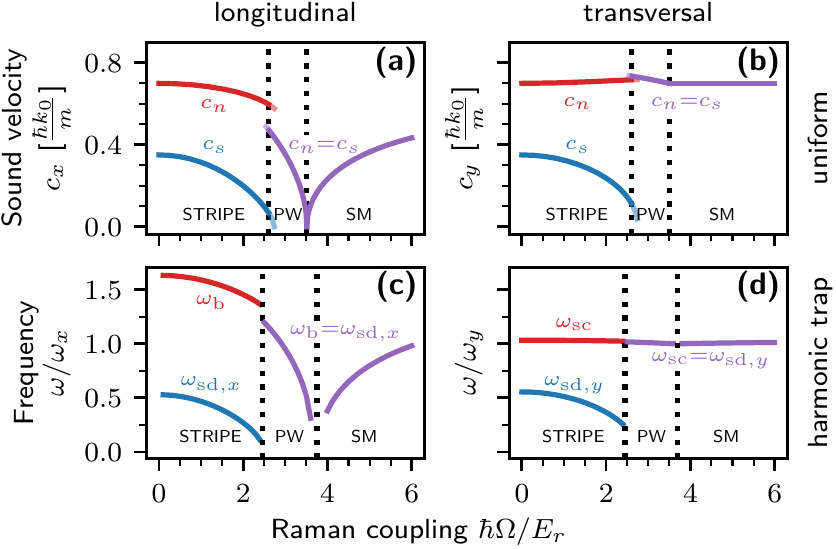}%
 	\subfloat{\label{fig:dispersion:a}}%
	\subfloat{\label{fig:dispersion:b}}%
	\subfloat{\label{fig:dispersion:c}}%
	\subfloat{\label{fig:dispersion:d}}%
	\caption{\label{fig:dispersion}%
        Sound velocities~$c$ in infinite uniform systems (a),(b) and frequencies~$\omega$ of collective excitations in finite-size harmonically trapped systems (c),(d) as a function of the Raman coupling~$\Omega$.
        The longitudinal~(a) and transversal~(b) sound velocities show the existence of two distinct sounds $c_{\densitycomponent}$ and $c_{\spincomponent}$ in the supersolid (stripe) phase, corresponding to the density and spin phonon branches, respectively.
        By contrast, there is only a single sound velocity of fully spin--density-hybridized nature in the plane-wave (PW)\protect\footnote{In the plane-wave phase, \cref{fig:dispersion:a} shows the geometric mean of the sound velocities corresponding to the positive and the negative part of the spectrum for a condensate with positive $x$~momentum, which is directly related to the magnetic susceptibility~\cite{Martone2012S} and the superfluid fraction~\cite{Martone2021Sc}.} and single-minimum (SM) phases.
        The light lines show the continuation of the sound velocity dispersion across the stripe--PW transition up to the respective spinodal points.
        (c)~Dispersion of the longitudinal breathing ($\omega_{\mathrm{b}}$) and spin-dipole mode ($\omega_{\mathrm{sd}, x}$) in a harmonic trap, excited by the operators $x^2$ and $x \sigma_z$.
        (d)~Dispersion of the scissors ($\omega_{\mathrm{sc}}$) and transversal spin-dipole mode ($\omega_{\mathrm{sd}, x}$) with corresponding excitation operators $x y$ and $y \sigma_z$.
        In the nonsupersolid phases, the density and spin operators excite the same collective modes.
        The dotted vertical lines mark the boundaries between the stripe, plane-wave, and single-minimum phases.%
	}
\end{figure}

As the strength of the Raman coupling~$\Omega$ increases, \acp{socbec} at moderate densities undergo a first-order transition between the supersolid stripe phase to the nonsupersolid plane-wave phase, followed by a second-order transition between the superfluid (but not supersolid) plane-wave and single-minimum phases~\cite{Lin2011S,Ho2011S,Li2012Sa}.
In \cref{fig:dispersion:a,fig:dispersion:b}, we show the dispersion of the sound velocities across the mean-field phase diagram, computed by numerically solving the linearized \ac{gp} equation in infinite systems.
In the stripe phase, below the critical Raman coupling~$\omegacritical$, two sounds $c_{\densitycomponent}$ and $c_{\spincomponent}$ of density and spin nature, respectively, are predicted to propagate due to the spontaneous breaking of both phase and translational invariance~\cite{Li2013S}.
At the transition between the stripe and plane-wave phase, the sound velocity~$c_{\spincomponent}$ of the spin Goldstone mode remains finite due to the first-order nature of the transition ($c_{\spincomponent}$ actually vanishes at the spinodal point, located at a slightly higher value of $\Omega$ than $\omegacritical$~\cite{Martone2021Sc}).
Since the relative phase of the two spin components is locked at $\Omega > \omegacritical$, the nonsupersolid phases feature only a single spin--density-hybridized sound, whose longitudinal velocity component vanishes at the second-order transition between the plane-wave and single-minimum phase due to the divergence of the effective mass~\cite{Martone2012S}.
The release of the locking of the relative phase in the supersolid phase at $\Omega < \omegacritical$ is the key physical effect enabling the excitation of the spin Goldstone mode and thus the dynamics of stripes, as explicitly revealed by the perturbation approach developed in Ref.~\cite{Martone2021Sc}.

Analytic expressions for the sound velocities in the supersolid phase are available up to second order in $\Omega$~\cite{Martone2021Sc}:
\begin{widetext}
\begin{subequations}
\label{eq:bogo2_sound_vel}
\begin{align}
\begin{split}
c_{\densitycomponent,x} = {}&{} c_\densitycomponent^{(0)}
- \frac{\left[ G_{\densitycomponent\densitycomponent}^3 \recoilenergy + 6 G_{\densitycomponent\densitycomponent}^2 \recoilenergy^2 + 2 G_{\densitycomponent\densitycomponent} \recoilenergy^2 \left(8\recoilenergy+G_{\spincomponent\spincomponent}\right)
+ 8 \recoilenergy^4 \right] c_\densitycomponent^{(0)}}{2 (2\recoilenergy+G_{\densitycomponent\densitycomponent})^3 (2\recoilenergy+G_{\spincomponent\spincomponent})}
\left(\frac{\hbar\Omega}{4 \recoilenergy}\right)^2 \, ,
\end{split}
\label{eq:bogo2_cx_n} \\
c_{\densitycomponent,\perp} = {}&{} c_\densitycomponent^{(0)} +
\frac{2 \recoilenergy^3 c_\densitycomponent^{(0)}}{(2\recoilenergy+G_{\densitycomponent\densitycomponent})^3}
\left(\frac{\hbar\Omega}{4 \recoilenergy}\right)^2 \, ,
\label{eq:bogo2_cp_n} \\
\begin{split}
c_{\spincomponent,x} = {}&{} c_\spincomponent^{(0)}
- \frac{c_\spincomponent^{(0)}}{2 G_{\spincomponent\spincomponent} (2\recoilenergy+G_{\densitycomponent\densitycomponent})^3 (2\recoilenergy+G_{\spincomponent\spincomponent})}
\big[ G_{\densitycomponent\densitycomponent}^3 \left(2\recoilenergy^2+5G_{\spincomponent\spincomponent}\recoilenergy+2 G_{\spincomponent\spincomponent}^2\right)
+ G_{\densitycomponent\densitycomponent}^2 \recoilenergy \left(8\recoilenergy^2+30G_{\spincomponent\spincomponent}\recoilenergy+13 G_{\spincomponent\spincomponent}^2\right) \\
&{} + G_{\densitycomponent\densitycomponent} \recoilenergy \left(8\recoilenergy^3+40G_{\spincomponent\spincomponent}\recoilenergy^2+22G_{\spincomponent\spincomponent}^2\recoilenergy+G_{\spincomponent\spincomponent}^3\right)
+ 2 G_{\spincomponent\spincomponent} \recoilenergy^2 \left(16\recoilenergy^2+12G_{\spincomponent\spincomponent}\recoilenergy+G_{\spincomponent\spincomponent}^2\right) \big]
\left(\frac{\hbar\Omega}{4 \recoilenergy}\right)^2 \, ,
\end{split}
\label{eq:bogo2_cx_s} \\
\begin{split}
c_{\spincomponent,\perp} = {}&{} c_\spincomponent^{(0)}
- \frac{c_\spincomponent^{(0)}}{2 G_{\spincomponent\spincomponent} (2\recoilenergy+G_{\densitycomponent\densitycomponent})^2 (2\recoilenergy+G_{\spincomponent\spincomponent})}
\big[ G_{\densitycomponent\densitycomponent}^2 \recoilenergy \left(2\recoilenergy+G_{\spincomponent\spincomponent}\right)
+ 2 G_{\densitycomponent\densitycomponent} \recoilenergy \left(2\recoilenergy^2+5G_{\spincomponent\spincomponent}\recoilenergy+2G_{\spincomponent\spincomponent}^2\right) \\
&{} + G_{\spincomponent\spincomponent} \recoilenergy \left(8\recoilenergy^2+8G_{\spincomponent\spincomponent}\recoilenergy+G_{\spincomponent\spincomponent}^2\right) \big]
\left(\frac{\hbar\Omega}{4 \recoilenergy}\right)^2 \, .
\end{split}
\label{eq:bogo2_cp_s}
\end{align}
\end{subequations}
\end{widetext}
Here, $c_\densitycomponent^{(0)} = \sqrt{G_{\densitycomponent\densitycomponent} / m}$ and $c_\spincomponent^{(0)} = \sqrt{G_{\spincomponent\spincomponent} / m}$ with $G_{\densitycomponent\densitycomponent} = \bar{n} g_{\densitycomponent\densitycomponent}$ and $G_{\spincomponent\spincomponent} = \bar{n} g_{\spincomponent\spincomponent}$ are the density and spin sound velocity at zero Raman coupling, respectively.
We employ the subscripts $x$ and $\perp$ to distinguish between the velocities of sound waves propagating along and perpendicular to the $x$~direction, which are different because of the anisotropic character of the \ac{soc} Hamiltonian.

A similar scenario also occurs for discretized collective modes in finite-size harmonically trapped systems.
In the nonsupersolid phases, above $\omegacritical$, the density and spin degrees of freedom are fully hybridized due to the locking of the relative phase.
Hence, both density and spin perturbations can excite the same Goldstone-like collective modes.
As an example, we recall that since the mechanical momentum operator is $\vect{p} - \hbar \vect{k}_0 \sigma_z$, a $\sigma_z$ perturbation creates momentum along $x$, and therefore both the $x$ and the $\sigma_z$ operator excite the center-of-mass (dipole) mode~\cite{Li2012S}. Similarly, the axial monopole (breathing or compression) mode is excited by both the operators $x^2$ and $x\sigma_z$~\cite{Geier2021S}, and the quadrupole (scissors) mode by both $xy$ and $y \sigma_z$ (see Ref.~\cite{Martone2012S}).
By contrast, in the supersolid phase, the above density and spin operators predominantly excite distinct collective modes. This is illustrated in \cref{fig:dispersion:c} for the longitudinal operators $x^2$ and $x \sigma_z$~\cite{Chen2017S,Geier2021S} as well as in \cref{fig:dispersion:d} for the transversal operators $x y$ and $y \sigma_z$, computed by numerically solving the full \ac{gp} equations in a harmonic trap using the same parameters as given in the main text.
The emergence of an additional Goldstone mode of spin nature below $\omegacritical$ leads to beating effects (see \cref{sec:beat_dens}), which have been proposed in Ref.~\cite{Geier2021S} as experimentally accessible signatures of supersolidity.

\section{Details on the density profile in infinite systems}
\label{sec:pert_app}

As pointed out in Ref.~\cite{Li2013S}, in the supersolid phase the condensate order parameter and the Bogoliubov
amplitudes have the form of Bloch waves. Starting from this result, one can easily prove that, after the release of an external perturbation transferring momentum $\hbar\vect{q}$, the total density exhibits the structure
\begin{equation}
\begin{split}
n(\vect{r},t) = {}&{} \frac{1}{2} \sum_{\bar{m} \in \mathbb{Z}}
\big[ \tilde{n}_{\bar{m}}
+ \delta \tilde{n}_{\vect{q},\bar{m}}(t) e^{i \vect{q} \cdot \vect{r}} \\
&{} + \delta \tilde{n}_{-\vect{q},\bar{m}}(t) e^{- i \vect{q} \cdot \vect{r}} \big] e^{i \bar{m} (2 k_1 x + \phi)} \, .
\end{split}
\label{eq:tot_dens_gen}
\end{equation}
Here, the coefficients of the equilibrium part are real and obey $\tilde{n}_{-\bar{m}} = \tilde{n}_{\bar{m}}$ and $\tilde{n}_{0} = 2 \bar{n}$, while the fluctuation terms fulfill the property $\delta \tilde{n}_{-\vec{q},\bar{m}}(t) = \delta \tilde{n}_{\vec{q},-\bar{m}}^*(t)$. At zero magnetic detuning ($\delta = 0$) and
symmetric intraspecies coupling ($g_{\uparrow\uparrow} = g_{\downarrow\downarrow}$), one can show that $\delta \tilde{n}_{\vec{q},\bar{m}}(t)$
is either real (for a density perturbation) or imaginary (for a spin perturbation). The time dependence of these coefficients is determined by the frequencies of the Bogoliubov spectrum, first studied in Ref.~\cite{Li2013S}.
In the case of a spin perturbation in the $q \ll m c_{\densitycomponent,\spincomponent} / \hbar, k_1$ limit, after taking the leading order in $q$ and neglecting the contributions of the gapped branches of the spectrum, \cref{eq:tot_dens_gen} reduces to Eq.~(3) in the main text.

In equilibrium, the density modulations exhibited by the supersolid phase can be characterized by their wave vector $2 \vect{k}_1 = 2 k_1 \hat{\vect{e}}_x$ and contrast $\mathcal{C} = (n_\mathrm{max} - n_\mathrm{min}) / (n_\mathrm{max} + n_\mathrm{min})$, with $n_\mathrm{min}$ ($n_\mathrm{max}$) the minimum (maximum)
value of the density. Up to second order in $\Omega$, these quantities are given by~\cite{Martone2021Sc}
\begin{align}
k_1 &{} = k_0- k_0 \frac{4 \recoilenergy^2 + 2 G_{\densitycomponent\densitycomponent} \recoilenergy + G_{\densitycomponent\densitycomponent}^2}{2(2\recoilenergy+G_{\densitycomponent\densitycomponent})^2}
\left( \frac{\hbar \Omega}{4 \recoilenergy} \right)^2 \, ,
\label{eq:gs2_k1} \\
\mathcal{C} &{} = \frac{2 \recoilenergy}{2 \recoilenergy + G_{\densitycomponent\densitycomponent}} \frac{\hbar \Omega}{4 \recoilenergy} \, .
\label{eq:gs1_contrast}
\end{align}

The behavior of the stripe wave vector with increasing Raman coupling can intuitively be understood in the laboratory frame as follows.
In the limit of weak \ac{soc}, the stripe pattern in the supersolid phase emerges from the spatial interference of condensates at zero momentum with their Raman sidebands at nonzero momenta within each spin component (see Fig.~1 in the main text, where for the present discussion we consider the case without additional spin current, i.e., $k_{\mathrm{rel}} = 0$).
For stronger \ac{soc}, kinetic energy is reduced when the two condensates are not at zero momentum, but at $\mp \hbar (k_0 - k_1)$ with \ac{soc} momentum sidebands at $\mp \hbar (k_0 + k_1)$ (here, the upper and lower sign refers to the up and down spin component, respectively).
This reduces the wave vector of the stripes from $2 k_0$ to $2 k_1$, as described by the perturbative expression in \cref{eq:gs2_k1}.
For very strong \ac{soc}, in the single-minimum phase, the condensate and its Raman sideband instead minimize the kinetic energy if they are at the momenta $\mp \hbar k_0$ and $k_1$ is zero.

It is also worth noticing that the spin perturbation affects the density profile not only at the microscopic level of stripe dynamics, but also at more macroscopic scales. This is revealed by the small-$q$ behavior of the $\vect{q}$-Fourier transform of the total density~\eqref{eq:tot_dens_gen}, providing the density response function to a spin perturbation. Indeed, due to \ac{soc}, a spin perturbation generally produces a weak cross excitation in the density sector at higher orders in $\Omega$, which leads to the occurrence of beat notes involving the frequencies of both the density and spin mode. One can additionally prove that an analogous effect takes place after the release of a density perturbation, inducing beat notes in the Fourier transform of the spin density at small $q$~\cite{Martone2023Sprep}. This cross-excitation mechanism is also responsible for the beat note in the stripe dynamics caused by a density perturbation, as discussed in \cref{sec:beat_dens}.

\section{Motion of stripes generated by uniform spin perturbations}
\label{sec:stripe_mot}

\begin{figure}[t]
 	\subfloat{\label{fig:velocity_stripe:a}}%
 	\subfloat{\label{fig:velocity_stripe:b}}%
 	\includegraphics[width=\columnwidth]{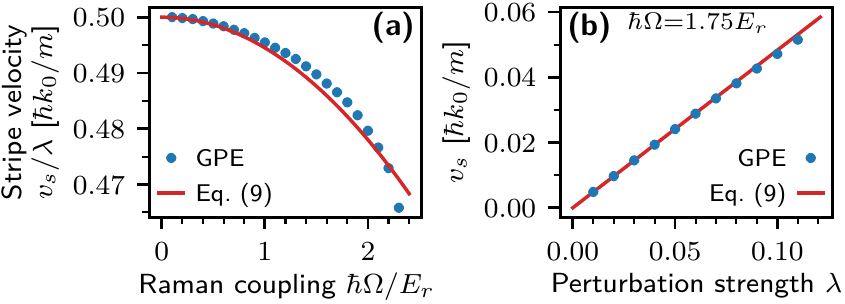}
 	\caption{\label{fig:velocity_stripe}%
        Velocity~$v_{\spincomponent}$ of the translational motion of the stripes (zero-frequency crystal Goldstone mode), excited by suddenly releasing the uniform spin perturbation $H_{\sigma_z} = - \lambda \recoilenergy \sigma_z$ in a harmonically trapped system.
        (a)~Stripe velocity as a function of the Raman coupling~$\Omega$ for a fixed perturbation strength $\lambda = \num{0.02}$.
        (b)~Stripe velocity as a function of the perturbation strength~$\lambda$ for a fixed Raman coupling $\hbar \Omega / \recoilenergy = \num{1.75}$.
        The numerical simulations based on the \acl{gp} equation (GPE) are in good agreement with the analytical prediction by Eq.~(9).%
    }
\end{figure}

In Ref.~\cite{Geier2021S}, it has been shown numerically that suddenly releasing a uniform spin perturbation by the operator~$\sigma_z$ excites the translational motion of the stripes, corresponding to the zero-frequency crystal Goldstone mode.
Intuitively, this can be understood as follows.
The spin perturbation in form of a Raman detuning creates a population imbalance with respect to equilibrium.
After the perturbation is switched off, there is a chemical potential difference between the two spin components.
Consequently, the relative phase of the two spin states, which is closely connected to the stripe wave vector through Eq.~(4), evolves linearly in time, causing a translation of the stripe pattern at a constant velocity~$v_{\spincomponent}$.
By contrast, no such effect can be observed using a uniform density perturbation.

In Ref.~\cite{Martone2021Sc}, it has been understood analytically within a perturbation approach in infinite uniform systems that the stripe motion is mainly associated with the behavior of the spin mode in the long-wavelength limit (the density mode for $q \to 0$ contributes to the stripe motion only at higher orders in $\Omega$, while at small Raman coupling it simply describes a shift of the condensate global phase).
In the present work, the velocity~$v_{\spincomponent}$ of the stripes has been calculated up to second order in $\Omega$ and the result is given by Eq.~(9).
We now examine with the help of numerical \ac{gp} simulations (for the same parameters as given in the main text) to what extent this expression also holds in finite-size harmonically trapped systems.

In \cref{fig:velocity_stripe}, we investigate the dependence of the stripe velocity~$v_{\spincomponent}$ on the Raman coupling~$\Omega$ as well as on the perturbation strength~$\lambda$ after suddenly releasing a uniform spin perturbation by the operator~$\sigma_z$.
The phase velocity of the stripes has been extracted from the numerical data as follows.
Phenomenologically, the time-dependence of the total density is well described by a \ac{tf} profile that is modulated by a traveling plane wave,
\begin{equation}
    \label{eq:density_tf}
    n(\vect{r}, t) = n_{\mathrm{TF}}(\vect{r}) \left\{ 1 + \mathcal{C} \cos \left[ 2 k_1 (x - v_{\spincomponent} t) + \phi \right] \right\} \,.
\end{equation}
Here, $n_{\mathrm{TF}}(\vect{r}) = [\mu_{\mathrm{TF}} - V(\vect{r})] / g$ [$n_{\mathrm{TF}}(\vect{r}) = 0$] inside [outside] the region $V(\vect{r}) \le \mu_{\mathrm{TF}}$ denotes the standard \ac{tf} density profile with chemical potential~$\mu_{\mathrm{TF}}$ and interaction constant~$g$~\cite{Pitaevskii2016S}, $\mathcal{C}$ is the contrast of the stripe pattern, $k_1$ the stripe wave number, and $v_{\spincomponent}$ the phase velocity of the stripes.
Note that in equilibrium, the origin of the stripe pattern in presence of a harmonic trap is fixed by energy minimization, yielding $\phi = 0$~\cite{Li2012Sa}.
The desired quantity~$v_{\spincomponent}$ can then be extracted from two fits of \cref{eq:density_tf} to the numerical data at fixed time $t = 0$ and fixed position $\vect{r} = 0$, respectively.

In order to compare the numerical results in \cref{fig:velocity_stripe} to the theoretical prediction, we have replaced the uniform density~$\bar{n}$ in Eq.~(9) by the mean central density of the cloud, i.e., the density averaged over the stripe period around the center of the trap.
This quantity can conveniently be obtained as $n_{\mathrm{TF}}(\vect{r} = 0)$ by fitting \cref{eq:density_tf} or it can be estimated \textit{a priori} using the \ac{tf} approximation for small values of the Raman coupling~$\Omega$.
Here, we follow the latter approach, which has the advantage of not requiring any free parameters.
To this end, one can show by varying the energy functional in Eq.~(2) that in uniform systems the spin density~$s_z$ vanishes for $\delta = 0$ and symmetric intraspecies interactions ($g_{\densitycomponent \spincomponent} = 0$) as $\Omega \to 0$.
Consequently, the total density in this limit is well approximated by the \ac{tf} profile of a single-component \ac{bec} with interaction constant~$g_{\densitycomponent \densitycomponent}$.
The \ac{tf} prediction for the central density obtained this way also provides a good approximation of the mean central density at stronger Raman couplings.

\Cref{fig:velocity_stripe:a} shows that the numerically extracted stripe velocity~$v_{\spincomponent}$ is in excellent agreement with the prediction by Eq.~(9) at small Raman couplings, while at higher values of $\Omega$ the deviation is at most $\SI{1}{\percent}$, even close to the phase transition.
Furthermore, \cref{fig:velocity_stripe:b} confirms the linear scaling of the stripe velocity with the perturbation strength~$\lambda$ at a fixed value of the Raman coupling, as predicted by Eq.~(9).
Linearity holds for a wide range of $\lambda$, up to the point when the perturbation becomes so strong that the stripe phase ceases to exist as the system becomes fully polarized.

\section{Beating effects caused by density perturbations}
\label{sec:beat_dens}

\begin{figure}
	\subfloat{\label{fig:beating:a}}%
	\subfloat{\label{fig:beating:b}}%
	\subfloat{\label{fig:beating:c}}%
	\subfloat{\label{fig:beating:d}}%
	\includegraphics[width=\columnwidth]{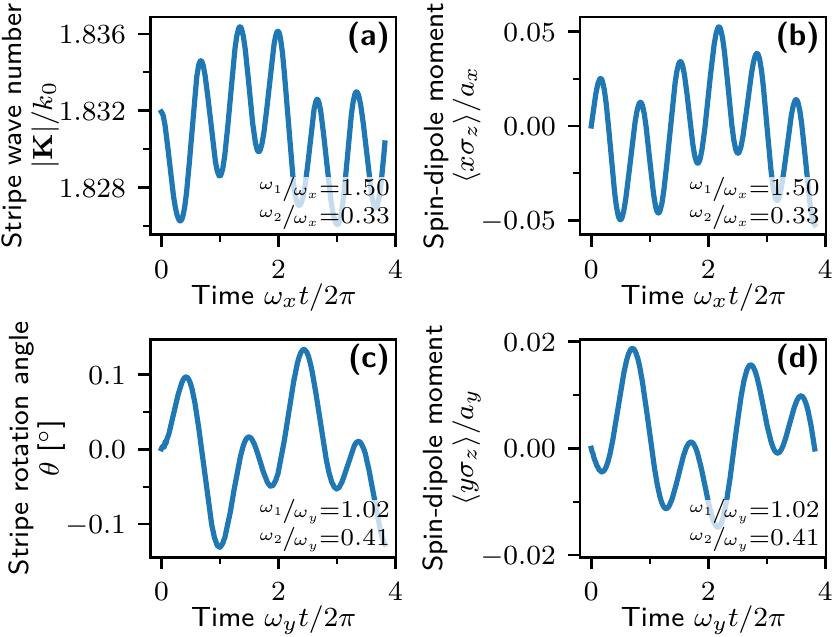}
	\caption{\label{fig:beating}%
        Beating effects in the stripe dynamics and spin-dipole oscillations caused by density perturbations in harmonically trapped systems for $\hbar \Omega / \recoilenergy = \num{1.75}$.
        (a)~Oscillations of the stripe wave vector~$\vect{K}$, whose magnitude determines the fringe spacing, and (b)~of the longitudinal spin-dipole moment~$\braket{x \sigma_z}$ after suddenly releasing the longitudinal density perturbation $H_{x^2} = \lambda m \omega_x^2 \mathinner{x^2}$ with $\lambda = 0.1$.
        Both observables exhibit a beat note involving the frequencies of the axial breathing mode ($\omega_1 / \omega_x = \num{1.50}$) and of the longitudinal spin-dipole mode ($\omega_2 / \omega_x = \num{0.33}$), cf.\ \cref{fig:dispersion:c}.
        (c)~Time evolution of the stripe rotation angle~$\theta$ and (d)~of the transversal spin-dipole moment~$\braket{y \sigma_z}$ following the sudden removal of the quadrupole perturbation $H_{xy} = -\lambda m \omega_x \omega_y \mathinner{x y}$ with $\lambda = 0.2$.
        Both signals feature a beating oscillation of two frequencies corresponding to the those of the scissors mode ($\omega_1 / \omega_y = \num{1.02}$) and of the transversal spin-dipole mode ($\omega_2 / \omega_y = \num{0.41}$), cf.\ \cref{fig:dispersion:d}.%
 	}
\end{figure}

As mentioned in the main text, stripe dynamics can also be excited by density perturbations as a consequence of \ac{soc}, which is accompanied by the occurrence of beat notes in the spin sector.
In this section, we provide an intuitive explanation of the mechanism behind such beating effects and illustrate this phenomenon with the help of numerical \ac{gp} simulations conducted in harmonically trapped systems (parameters as given in the main text).
For a quantitative discussion in infinite systems, we refer the interested reader to Ref.~\cite{Martone2023Sprep}.

As a basic remark, the fact that the wave number~$k_1$ in \cref{eq:gs2_k1}, fixing the relative distance between stripes, depends on the density~\cite{Li2012Sa,Martone2021Sc} suggests that a modulation of the density induces a variation of the fringe spacing.
The dependence of the stripe spacing on the \ac{soc} strength is weaker at high densities since the mean-field energy increases the effective detuning of the Raman coupling.
As a result, an oscillation in the density will then lead to an oscillation of the fringe spacing.

Going beyond this basic picture, the coupling between the spin and density character of the two Goldstone modes typically leads to beating effects when either a spin or a density perturbation is applied.

At small Raman couplings~$\Omega$, a spin perturbation creates a strong excitation of the spin Goldstone mode, which affects the microscopic density (i.e., the density at length scales on the order of the fringe spacing, reflecting the behavior of the relative phase of the two spin components), but not the macroscopic density (i.e., the density coarse-grained over typical length scales on the order of the fringe spacing).
Likewise, a density perturbation mainly drives the density Goldstone mode.

At higher values of $\Omega$ (still below $\omegacritical$), nonlinear effects in the \ac{soc} strength become manifest in a mixing of spin and density degrees of freedom.
As a result, a spin perturbation also weakly excites the density Goldstone mode.
Since the spin Goldstone mode has a weak component in the density sector, there is now a beat note between the two Goldstone modes, visible in the macroscopic density.
It can be shown within the perturbation approach in infinite systems described in the main text that the amplitude of this beat note is of order $\Omega^2$~\cite{Martone2023Sprep}.
By contrast, the beat note in the dynamics of the stripe wave vector is almost invisible [see Fig.~2 in the main text] since the density mode is only weakly excited and its contribution to the relative phase is of order $\Omega^4$.
Vice versa, a density perturbation also creates a weak response in the spin sector and therefore in the stripe dynamics.
Since the density mode has a weak spin component, there is now a beat note between the two Goldstone modes, visible in the spin sector and therefore in the fringe dynamics.

\Cref{fig:beating} shows the beating oscillations in the spin sector induced by a longitudinal (transversal) density perturbation proportional to the operator $x^2$ ($x y$) in a harmonically trapped system.
In the longitudinal case, \cref{fig:beating:a,fig:beating:b}, the magnitude of the stripe wave vector~$\abs{\vect{K}}$ (and thus the fringe spacing) as well as the longitudinal spin-dipole moment~$\braket{x \sigma_z}$ exhibit a beat note involving the frequencies of the density breathing mode and the longitudinal spin-dipole mode, whose dispersion relation is shown in \cref{fig:dispersion:c}.
As discussed in \cref{sec:disp_coll_osc}, these two modes, which are mainly excited by the density operator~$x^2$ and the spin operator~$x \sigma_z$, respectively, are fully hybridized in the superfluid phases for $\Omega > \omegacritical$, while in the supersolid phase for $\Omega < \omegacritical$ the partial spin--density hybridization explained above is responsible for the observed beating effect.
The amplitude of the fringe spacing oscillation excited by the density operator~$x^2$ is of order $\Omega^2$ and thus smaller than in case of a direct excitation of the relevant spin mode by the spin operator~$x \sigma_z$, where no beating is visible [cf.\ Fig.~2(c) in the main text].

Analogously, in the transversal case, \cref{fig:beating:c,fig:beating:d}, the orientation of the stripes as well as the transversal spin-dipole moment~$\braket{y \sigma_z}$ undergo beating oscillations involving the frequencies of the density scissors mode and the transversal spin-dipole mode.
These modes, which are mainly excited by the density operator~$xy$ and the spin operator~$y \sigma_z$, hybridize under \ac{soc} and their dispersion law is shown in \cref{fig:dispersion:d}.

The emergence of beat notes between the density and spin Goldstone modes underlines the crucial role of the coupling between density and spin degrees of freedom in \ac{soc} configurations and opens up new possibilities for the dynamic excitation of the supersolid stripe pattern, also with regard to future experiments.

\end{document}